\documentstyle[12pt]{article}
\begin{document}
 \bibliographystyle{unsrt}
\begin{center}
{\Large {\bf 
Photon-number tomography of multimode states and 
positivity of the density matrix}}
\end{center}

\begin{center}
Olga~ Man'ko and  V.~ I.~ Man'ko 
\end{center}

\begin{center}
P.\,N.\,Lebedev Physical Institute,
Leninskii Prospect 53, Moscow 119991, Russia\\
e-mail: mankoov@lebedev.ru; mankovi@lebedev.ru
\end{center}

\begin{center}
{\bf  Abstract}\\
\end{center}
For one-mode and multimode light, the photon-number tomograms of 
Gaussian quantum states are explicitly calculated in terms of multivariable 
Hermite polynomials. Positivity of the tomograms is shown to be 
necessary condition for positivity of the density matrix.

\section{Introduction}
\noindent The photon-number tomography was introduced in ~[1--3].
The name ``photon-number tomography'' was first used
in~\cite{euroPLMancini} due to the physical meaning of measuring
the quantum state by measuring number of photons (i.e., photon
statistics). Photon-number tomography is the method to reconstruct the
density operator of a quantum state employing measurable probability
distribution function (photon statistics) called tomogram.
Photon-number tomography differs of the optical tomography
method~\cite{bertrand,vogelrisken} and the symplectic
tomography scheme~[6--8]
where continuous homodyne quadratures are measured for reconstructing
the quantum state. In photon-number tomography, a discrete random
variable is measured for reconstructing quantum state (photon density matrix).
The other tomography, where probability distributions of discrete random
variables are used, is spin tomography~[9--15].
In spin tomography, the discrete random variables (spin projections) vary 
in the finite domain $-j\leq m\leq j$. In photon-number tomography,
the discrete random variables (number of photons) vary in infinite
domain $0 \leq n \leq \infty$.

The aim of the paper is to discuss properties of photon-number tomograms
for Gaussian quantum states in the one-mode and multimode cases and
find a criterion of positivity of the density matrix.

The paper is organized as follows.

In Sec.~2 we review properties of the photon-number tomograms while
in Sec.~3 we discuss the photon-number tomograms of multimode light.
Positivity of the density matrix and its relation to positivity of the
photon-state tomographic symbol is studied in Sec.~4. Conclusions and
perspectives of the approach are given in Sec.~5.

\section{One-Mode Case}
The photon-number tomogram defined by the relation
\begin{equation}\label{eq.1}
\omega(n,\alpha)=\langle n\mid\hat D(\alpha)\hat
\rho\hat D^{-1}(\alpha)\mid n\rangle
\end{equation}
is the function of integer photon number $n$ and complex number 
$$\alpha=\mbox{Re}\,\alpha+i\,\mbox{Im}\,\alpha,$$ 
where $\hat\rho$ is the state density operator and
$\hat D(\alpha)$ is the Weyl displacement operator 
$$\hat D(\alpha)=\exp(\alpha\hat a^{\dagger}-\alpha^*\hat a).$$ 
It is known~\cite{cahillglauber} that the Wigner function,
which corresponds to the density operator $\hat\rho$, is given
by the expression
\begin{equation}\label{eq.2}
W_{\hat\rho}(q,p)=
2\,\mbox{Tr}\left[\hat\rho\hat D(\beta)
\left(-1\right)^{\hat a^{\dagger}\hat a}\hat D(-\beta)\right],
\end{equation}
where $\hat D(\beta)$ is the Weyl displacement operator with complex argument 
$$\beta=\frac{1}{\sqrt2}(q+ip), $$
with
$\hat a$ and $\hat a^{\dagger}$ being
photon annihilation and creation operators.

Let us introduce the displaced density operator 
\begin{equation}\label{eq.3}
\hat\rho_{\alpha}=\hat D^{-1}(\alpha)\hat\rho\hat D(\alpha).
\end{equation}
The Wigner function, which corresponds to the displaced density operator,
is of the form 
\begin{equation}\label{eq.4}
W_{\hat\rho_{\alpha}}(q,p)=
2\,\mbox{Tr}\left[\hat\rho_{\alpha}\hat D(\beta)\left(-1\right)^{\hat a^{\dagger}\hat a}\hat D(-\beta)\right].
\end{equation}
By inserting the expression for the displaced density operator into
(\ref{eq.4}), one arrives at
\begin{equation}\label{eq.5}
W_{\hat\rho_{\alpha}}(q,p)=
2\,\mbox{Tr}\left[\hat D^{-1}(\alpha)\hat\rho\hat D(\alpha)\hat D(\beta)
\left(-1\right)^{\hat a^{\dagger}\hat a}\hat D(-\beta)\right].
\end{equation}
In view of the properties of the Weyl displacement operator 
$$\hat D(\beta)\hat D(\alpha)=\hat D(\beta+\alpha)\exp\Big[i\,
\mbox{Im}\,(\beta^*\alpha)\Big],$$
$$\hat D^{-1}(\alpha)=\hat D(-\alpha),$$
$$\hat D^{-1}(\alpha)\hat D^{-1}(\beta)=
\Big(\hat D(\beta)\hat D(\alpha)\Big)^{-1},$$
formula~(\ref{eq.5}) can be simplified 
\begin{equation}\label{eq.6}
W_{\hat\rho_{\alpha}}(q,p)=
W_{\hat\rho}\Big(q+\sqrt{2}\,\mbox{Re}\,\alpha, ~p
+\sqrt{2}\,\mbox{Im}\,\alpha\Big).
\end{equation}
One can see that the Wigner function~(\ref{eq.4}) corresponding to the displaced density 
operator is equal to the Wigner function~(\ref{eq.2}) corresponding to the initial density 
operator but with displaced arguments. 

The photon-number tomogram is the photon distribution function (the 
probability to have $n$ photons) in the state described by the displaced density 
operator $\hat\rho_{\alpha}$~(\ref{eq.3}), i.e., 
\begin{equation}\label{eq.7}
\omega(n,\alpha)=P_n(\alpha)=\langle n |\rho_{\alpha}| n \rangle,\qquad
n=0,1,2,\ldots
\end{equation}
The photon distribution function for one-mode mixed light, 
described by the Wigner function of generic Gaussian form  
\begin{equation}\label{eq.8}
W(q,p)=\frac{1}{\sqrt{\mbox{det}\,\sigma(t)}}
\exp \left(-\frac {1}{2}
\mbox{\bf Q}\sigma^{-1}(t)\mbox{\bf Q}^T\right), 
\end{equation} 
where $\mbox{\bf Q}=\Big(p-\langle p\rangle,q-\langle q\rangle\Big)$ and
the matrix $\sigma(t)$ is real symmetric quadrature variance matrix 
$$\sigma(t)=\pmatrix{\sigma_{p p}&\sigma_{p q}\cr\sigma_{p q}&\sigma_{q q}},$$
was obtained explicitly in terms of the Hermite polynomials of two variables
in [17]. The quadature means and dispersions in the above formulas can depend
on time.

The Hermite polynomials of two variables $H^{\{\mbox {\bf R}\}}_{n_1\,n_2}(y_1,y_2)$, 
where $n_1,n_2$ are nonnegative integers and $\mbox{\bf R}$
is a symmetric 2$\times$2 matrix, are determined by the generating function 
\begin{equation}\label{eq.10} 
\exp\left[-{1\over2}(x_1\,x_2)\pmatrix{R_{11}&R_{12}\cr R_{21}&R_{22}}
\pmatrix{x_1\cr x_2}+(y_1\,y_2)\pmatrix{R_{11}&R_{12}\cr R_{21}&R_{22}}
\pmatrix{x_1\cr x_2}\right]=\sum_{n_1,n_2=0}^{\infty}\frac{x_1^{n_1}x_2^{n_2}}{n_1!n_2!}
\,H^{\{\mbox {\bf R}\}}_{n_1\, n_2}(y_1,y_2). 
\end{equation}
The photon distribution functions of nonclassical states of 
light described by the Gaussian wave functions were discussed 
in~[18--20].

Applying the scheme of calculations similar to the one used in \cite{21} 
to our photon-number tomogram~(\ref{eq.7}), we arrive at the photon-number 
tomogram as a function of the Hermite polynomial of two variables 
\begin{equation}\label{eq.11}
\omega(n,\alpha)=\frac{P_0(\alpha)H^{\{\mbox {\bf R}\}}_{n\,n}
\Big(y_1(\alpha),y_2(\alpha)\Big)}{n!}, 
\end{equation}
where the matrix $\mbox {\bf R}$, which determines the Hermite polynomial, reads  
$$\mbox{\bf R}=\frac{1}{1+2T+4d}
\pmatrix{2\left(\sigma_{pp}-\sigma_{qq}-2i\sigma_{pq}\right)
&1-4d \cr 1-4d&2\left(\sigma_{pp}-\sigma_{qq}+2i\sigma_{pq}\right)}.$$
Here $d$ is the determinant of real symmetric quadrature variance matrix 
$\sigma(t)$, i.e.,
$$d=\sigma_{pp}\sigma_{qq}-\sigma_{pq}^2$$
and $T$ is its trace 
$$T=\sigma_{pp}+\sigma_{qq}.$$
The arguments of the Hermite polynomial are 
\begin{eqnarray}
y_1(\alpha)=y_2^*({\alpha})&=&\frac{\sqrt{2}}{2T-4d-1}
\left[\left(\langle q\rangle-i\langle p\rangle+
\sqrt2\alpha^*\right)\left(T-1\right)\right.\nonumber\\
&&+\left.\left(\sigma_{pp}-\sigma_{qq}+2i\sigma_{pq}\right)\left(\langle q\rangle+
i\langle p\rangle+\sqrt2\,\alpha\right)\right].
\end{eqnarray}
For the state with displaced Wigner function~(\ref{eq.4}), the probability to have 
no photons $P_0(\alpha)$ reads 
\begin{eqnarray}
P_0(\alpha)&=&\frac{2}{\sqrt L}\exp\left\{-\frac{1}{L}\left[
\left(2\sigma_{qq}+1\right)\left(\langle p\rangle
+\sqrt{2}\,\mbox{Im}\,\alpha\right)^2
+\left(2\sigma_{pp}+1\right)\left(\langle q\rangle+\sqrt{2}\,
\mbox{Re}\,\alpha\right)^2
\right]\right\}
\nonumber\\
&&\times\exp\left[\frac{4\sigma_{p q}}{L}\left(\langle p\rangle+
\sqrt{2}\,\mbox{Im}\,\alpha\right)\left(\langle q\rangle
+\sqrt{2}\,\mbox{Re}\,\alpha\right)\right],\label{eq.12a}
\end{eqnarray}
where $L=1+2T+4d$.

The density operator can be reconstructed from the photon-number tomogram 
with the help of the inverse formula~[1--3]
\begin{equation}\label{eq.12b}
\hat\rho=\sum_{n=0}^{\infty}\int \frac{4\,d^2\alpha}{\pi(1-s^2)}
\left(\frac{s-1}{s+1}\right)^{(\hat a^\dagger+\alpha^*)
(\hat a+\alpha)-n}\omega(n,\alpha),
\end{equation} 
where $s$ is an arbitrary ordering parameter~\cite{cahillglauber}. 

Within the framework of a well-known method of the star-product quantization
(see, e.g., \cite{marmo}) one has the following relations: \\
(i) The photon-number tomogram is a symbol of the density operator 
\begin{equation}\label{eq.12c}
\omega(n,\alpha)=\mbox{Tr}\left[\hat\rho\,\hat U(\mbox{\bf x})\right], 
\end{equation}
where operator $\hat U(\mbox{\bf x})$ reads
$$\hat U(\mbox{\bf x})=\hat D(\alpha)|n\rangle\langle n|\hat D^{-1}(\alpha),
\,\mbox{\bf x}=(n,\alpha);$$ 
(ii) The density operator can be also expressed through its symbol 
\begin{equation}\label{eq.13}
\hat\rho=\sum_{n=0}^{\infty}\int d^2\alpha\,\omega(\mbox{\bf x})\hat D(\mbox{\bf x}).
\end{equation}
Comparing~(\ref{eq.12b}) with~(\ref{eq.13}) one can see that
\begin{equation}\label{eq.14}
\hat D(\mbox{\bf x})=\frac{4}{\pi(1-s^2)}
\left(\frac{s-1}{s+1}\right)^{(\hat a^\dagger+\alpha^*)(\hat a+\alpha)-n}.
\end{equation}
Now we consider some simple cases.

If the electromagnetic field is in the coherent state $|\gamma\rangle$, the
photon-number tomogram reads
$$
\omega_{\gamma}(n,\alpha)=\frac{1}{n!}|\gamma+\alpha|^{2n}
\exp\left[-|\gamma+\alpha|^2\right].
$$
One has the photon-number tomogram for squeezed and correlated states of the 
form 
\begin{eqnarray}
\omega_{sq}(n,\alpha)&=&\frac{\tanh^nr}{n!\,2^n\cosh r}
\exp\left[\tanh r\sin \theta\Big(\langle p\rangle+\sqrt{2}\,\mbox{Im}
\,\alpha\Big)
\Big(\langle q\rangle+\sqrt{2}\,\mbox{Re}\,\alpha\Big)\right.\nonumber\\
&&-\frac{1}{2}\Big(\langle p\rangle+\sqrt{2}\,\mbox{Im}\,\alpha\Big)^2
\left(1-\cos\theta\tanh r\right)\\
&&-\left.\frac{1}{2}\Big(\langle q\rangle+\sqrt{2}\,\mbox{Re}\,\alpha\Big)^2
\left(1+\cos\theta\tanh r\right)\right]\nonumber\\
&&\times\left|H_n\left\{\frac{1}{2}e^{-i\theta/2}\sqrt{\tanh r}\Big[
\langle q\rangle-i\langle p\rangle+\sqrt{2}\,\alpha^* 
+e^{i\theta}\coth r\Big(\langle q\rangle+i\langle p\rangle
+\sqrt{2}\,\alpha\Big)\Big]\right\}\right|^2,\nonumber\\
&&\label{eq.14a}\end{eqnarray}
where
$$\sin\theta=\frac{2\,\sigma_{pq}}{\sqrt{(\sigma_{pp}+\sigma_{qq})^2-1}},\qquad
\,\cosh\,2r=T.$$
Thus for Gaussian states, we constructed tomograms, which are positive
probability distributions of number of photons.
The tomograms determine the density operator of the quantum state completely. 

\section{Multimode Case }
Now we briefly discuss the case of a multimode mixed state of the 
electromagnetic field. It will be a generalization of multiparticle spin
tomography with discrete random variables varying in the finite
domain~\cite{andreevsafonov} to the case of multimode photon-number
tomography with discrete random variables varying in infinite domain
$0\leq n_1\leq\infty,~0 \leq n_2 \leq \infty,\ldots,
~0 \leq n_N \leq \infty.$

In the case of multimode light, the tomogram reads
\begin{equation}\label{eq.15}
\omega(\mbox{\bf n},
\vec{\alpha})
=\langle\mbox{\bf n}|\hat D(\vec{\alpha})
\hat \rho\hat D^{-1}(\vec{\alpha})
|\mbox{\bf n}\rangle=
\langle\mbox{\bf n}|\hat\rho_{\alpha} |\mbox{\bf n}\rangle.
\end{equation}
Components of the vector $\mbox{\bf n}$ are the integer photon numbers 
in different modes
$$\mbox{\bf n}=\left(n_1,n_2,\ldots,n_N\right)$$
and components of the vector $\vec{\alpha}$ are complex numbers 
$$\vec{\alpha}=\left(\alpha_1,\alpha_2,\ldots,\alpha_N\right),$$
where
$$\alpha_k=\mbox{Re}\,\alpha_k+i\,\mbox{Im}\,\alpha_k,\qquad
k=1,\ldots,N.$$
The displacement operator $\hat D(\vec{\alpha})$ is a product of the Weyl 
displacement operators of each mode 
$$\hat D(\vec{\alpha})=\prod_{k=1}^N\hat D(\alpha_k)=
\prod_{k=1}^N \exp\Big(\alpha_k\hat a_k^{\dagger}-\alpha_k^*\hat a_k\Big).$$
The Wigner function, which corresponds to the density operator $\hat\rho$, is 
given by the expression 
\begin{equation}\label{eq.16}
W_{\hat\rho}(q_1,\ldots,q_N,p_1,\ldots,p_N)=
2^N\mbox{Tr}\left[\hat\rho\hat D(\vec{\beta})
\left(-1\right)^{\hat a_1^{\dagger}\hat a_1+\cdots +\hat a_N^{\dagger}\hat a_N}
\hat D^{-1}(\vec{\beta})\right], 
\end{equation}
where $\hat D(\vec{\beta})$ is a product of the Weyl displacement operators 
for each mode. The vector $\vec{\beta}$ has complex components 
$\beta_k$ $(k=1,\ldots ,N)$, i.e.,  
$$\vec{\beta}=\left(\beta_1,\ldots ,\beta_N\right).$$
The Weyl displacement operator, which corresponds to the $k$th mode, 
$\hat D(\beta_k)$ has complex argument 
$$\beta_k=\frac{1}{\sqrt 2}(q_k+ip_k),$$
and it reads
$$\hat D(\beta_k)=e^{\beta_k\hat a_k^{\dagger}-\beta_k^*\hat a_k}. $$
Repeating the scheme of calculations used for the one-mode case, we obtain
\begin{eqnarray}
&&W_{\hat\rho_{\vec{\alpha}}}(q_1,\ldots ,q_N,p_1,\ldots ,p_N)\nonumber\\
&&=W_{\hat\rho}\Big(q_1+\sqrt{2}\,\mbox{Re}\,\alpha_1,\ldots ,
~q_N+\sqrt{2}\,\mbox{Re}\,\alpha_N,
~p_1+\sqrt{2}\,\mbox{Im}\,\alpha_1,\ldots ,~p_N+\sqrt{2}\,\mbox{Im}\,
\alpha_N\Big).
\end{eqnarray}
One can see that the Wigner function, which corresponds to the displaced 
density operator in the multimode case, is equal to the Wigner function 
corresponding to the initial density operator but with displaced arguments.

The photon distribution function of $N$-mode mixed state of 
light described by the Wigner function of a generic Gaussian form 
\begin{equation}\label{eq.18}
W(\mbox{\bf Q}')=\frac{1}{\sqrt{\mbox{det}\,\sigma(t)}}\,
\exp\left(-\frac{1}{2}\mbox{\bf Q}'
\sigma^{-1}(t)\mbox{\bf Q}'\right),
\end{equation}
where $2N$-dimensional vector $\mbox{\bf Q}'$ is
$$\mbox{\bf Q}'=\Big(p_1-\langle p_1\rangle,\ldots ,~p_N-\langle p_N\rangle,
~q_1-\langle q_1\rangle,\ldots ,~q_N-\langle q_N\rangle\Big)$$ 
and the matrix $\sigma(t)$ is  $2N$$\times$$2N$ real symmetric quadrature 
variance matrix, can be calculated explicitly in terms of the Hermite polynomials of 
$2N$ variables~\cite{23}.

The Hermite polynomial of $N$ variables
$ H^{\{\mbox{\bf R}\}}_{n_1\ldots n_N}(\mbox{\bf y})$
is determined by the generating function
$$ 
\exp\left(-{1\over2}\mbox{\bf x}\mbox{\bf R}\mbox{\bf x}^T+\mbox{x}R\mbox{\bf y}^T\right)
=\sum_{n_1,\ldots ,n_N=0}^{\infty}\frac{x_1^{n_1}}{n_1!}\cdots
\frac{x_N^{n_N}}{n_N!}
H^{\{\mbox{\bf R}\}}_{n_1\ldots n_N}(\mbox{\bf y}),
$$
where $n_k$ $(k=1,\ldots ,N)$ are nonegative integers and
$\mbox{\bf R}$ is a symmetric $N$$\times $$N$ matrix.

Applying the scheme of calculations of \cite{23} one can derive
the photon-number tomogram in the case of multimode
electromagnetic field as a function of the Hermite polynomial
of $2N$ variables
\begin{equation}\label{eq.19}
\omega(n_1,\ldots ,n_N,\alpha_1,\ldots ,\alpha_N)=\frac{P_0(\vec{\alpha})
H^{\{\mbox{\bf R}\}}_{n_1\ldots n_N\,n_1\ldots n_N}\Big(
\mbox{\bf y}(\vec{\alpha}),\mbox{\bf y}^*(\vec{\alpha})\Big)}
{n_1! \cdots n_N!}.
\end{equation}
The matrix $\mbox{\bf R}$, which determines the Hermite polynomial, reads
\begin{equation}\label{eq.19a}
\mbox{\bf R}=U^\dagger\Big(E_{2N}-2\sigma(t)\Big)\Big(E_{2N}
+2\sigma(t)\Big)^{-1}U^*,
\end{equation}
where $2N$$\times$$2N$ matrix $U$ consists of $N$-dimensional unity matrices 
$E_N$ with different coefficients 
$$U=\frac{1}{\sqrt2}\pmatrix{-i E_N & i E_N \cr E_N&E_N}$$ 
and $E_{2N}$ is $2N$-dimensional unity matrix. 
The argument of Hermite polynomial reads
\begin{equation}\label{eq.19b}
\mbox{\bf y}=2U^T\Big(E_{2N}-2\sigma(t)\Big)^{-1}(\mbox{\bf u},\mbox{\bf v})^T,
\end{equation}
where 
$$\mbox{\bf u}=\Big(\langle p_1\rangle+\sqrt2\,\mbox{Im}\,\alpha_1, \dots ,
\langle p_N\rangle+\sqrt2\,\mbox{Im}\,\alpha_N\Big)$$
and 
$$\mbox{\bf v}=\Big(\langle q_1\rangle+\sqrt2\,\mbox{Re}\,\alpha_1,
\ldots ,\langle q_N\rangle+\sqrt2\,\mbox{Re}\,\alpha_N\Big).$$
For a state with the Wigner function
$W_{\hat\rho}(q_1,\ldots ,q_N,p_1,\ldots ,p_N)$,
the probability to have no photons $P_0$ is given by the relation
\begin{equation}\label{eq.19c}
P_0=\frac{1}{\sqrt{\mbox {det}\,\Big(\sigma(t)+\frac{1}{2}E_{2N}\Big)}}
\exp\left[-(\mbox{\bf u},\mbox{\bf v})
\Big(2\sigma(t)+E_{2N}\Big)^{-1}
(\mbox{\bf u},\mbox{\bf v})^T\right].
\end{equation}
It can be proved that the density operator can be reconstructed from
the photon-number tomogram with the help  of the inversion formula
\begin{eqnarray}\label{A}
\hat\rho&=&\int\left[\prod_{k=1}^N \frac{d\,\mbox{Re}\,\alpha_k~ d\,\mbox{Im}
\,\alpha_k}{\pi}\right]\left\{\sum_{n_1=0}^\infty \cdots \sum_{n_N=0}^\infty
\left[\prod_{k'=1}^N\frac{2}{(1-s_{k'})}\left(\frac{s_{k'}+1}{s_{k'}-1}
\right)^{n_{k'}}\right]\right.\nonumber\\
&&\times\left.\omega(n_1,\ldots ,n_N,\alpha_1,\ldots ,\alpha_N)\hat T\right\},
\end{eqnarray}
where operator $\hat T$ reads
$$
\hat T=\prod_{k=1}^N\left[\frac{2}{1+s_k}\hat D^{-1}(\alpha_k)
\left(\frac{s_k-1}{s_k+1}\right)^{\hat a^\dagger_k\hat a_k}\hat D(\alpha_k)\right].
$$
Here $s_k$ are arbitrary ordering parameters~\cite{cahillglauber}. 
Employing the tomogram~(\ref{eq.15}) one can reconstruct a generic (squeezed,
correlated, and entangled) Gaussian density matrix of multimode light.

\section{Positivity of Density Matrix}
In this section, we discuss a criterion of positivity of the density matrix. 
We relate properties of tomographic symbols with positivity of
the Hermitian density matrix.
First, we remind the conditions of positivity of a density matrix.
Any Hermitian matrix $R$ (both finite or infinite dimensional one)
has real eigenvalues $R_k$.
It can be represented as a sum 
\begin{equation}\label{eq.s31}
R=\sum_k R_k|k\rangle\langle k|, 
\end{equation}
where $k$ is either discrete or continuous index
(for infinite-dimensional matrix). 
The projectors (or projector densities) $|k\rangle\langle k|$ satisfy the condition
\begin{equation}\label{eq.s32}
R|k\rangle \langle k|=R_k|k\rangle \langle k|.
\end{equation}
The nonnegative Hermitian operators satisfy the condition 
\begin{equation}\label{eq.s33}
R_k\geq 0.
\end{equation}
We formulate a linear criterion of nonnegativity of the Hermitian
operator $\hat\rho$. To do this, let us consider such operator as a density
operator of a physical system. One can associate with this operator
the tomographic symbol $\omega_{\hat\rho}$ of any kind
(optical tomographic~\cite{bertrand,vogelrisken,raymer}, symplectic 
tomographic~[6--8], 
photon-number tomographic~[1--3],
or spin-tomographic one~[9--15]).
Since the tomographic symbols of the density operators of quantum states have the 
physical meaning of the probability distribution (or the probability density
in the infinite
dimensional case), one has the inequality 
\begin{equation}\label{eq.s34}
\omega_{\hat\rho}\geq 0.
\end{equation}
This inequality is necessary (and in some cases sufficient) condition
of positivity of the density operator.

Let us consider this criterion for the photon-number tomograms. Since 
the photon-number tomogram completely determines the state, positivity
of the density operator is given by the explicit relation 
\begin{equation}\label{eq.s35}
\omega_{\hat\rho}(\mbox{\bf n},\vec{\alpha})\geq 0,
\end{equation}
where the tomogram $\omega_{\hat\rho}(\mbox{\bf n},\vec{\alpha})$ is determined in 
terms of the operator $\hat \rho$ as follows 
\begin{equation}\label{eq.s36}
\omega_{\hat\rho}(\mbox{\bf n},\vec{\alpha})=
\mbox{Tr}\left[\hat\rho\hat D(\vec{\alpha})|
\mbox{\bf n}\rangle\langle\mbox{\bf n}|\hat D(-\vec{\alpha})\right].
\end{equation}
If one takes an Hermitian operator $\hat R^{\dagger}=\hat R$,
which has some negative eigenvalues, one can get 
\begin{equation}\label{eq.s37}
\omega_{\hat R}(\mbox{\bf n},\vec{\alpha}) < 0
\end{equation}
for some values of $\mbox{\bf n}$ and $\vec{\alpha}$. 
This criterion can be expressed as a criterion for admissible
Wigner functions $W(\mbox{\bf q},\mbox{\bf p})$.
The Wigner function is admissible (i.e., it correspond to a state of
a quantum system), if one has 
\begin{equation}\label{eq.s38}
\omega_{\hat\rho}(\mbox{\bf n},\vec{\alpha})=
\int W\Big(\mbox{\bf q}+\mbox{Re}\,\vec{\alpha},\mbox{\bf p}
+\mbox{Im}\,\vec{\alpha}\Big)
\prod_{k=1}^N\left(2e^{-p_k^2-q_k^2} L_n\Big[2(p_k^2+q_k^2)\Big]
\right)\frac{d\,q_k~d\,p_k}{2\pi}\geq 0
\end{equation}
for any vector $\mbox{\bf n}=(n_1,n_2,\ldots , n_N)$ with $n_k=0,1,2,\ldots $ and arbitrary complex 
vector $\vec{\alpha}$. 
In the case of the Wigner function, which is the Weyl symbol of an
Hermitian (nonpositive) operator $\hat R$, the analogous integral can take
negative values, i.e.,
\begin{equation}\label{eq.s39}
\omega_{\hat R}(\mbox{\bf n},\vec{\alpha})< 0
\end{equation}
for some $\mbox{\bf n},\,\vec{\alpha}$. For Gaussian Hermitian operators, the criterion takes the form of inequality
\begin{equation}\label{eq.s310}
\frac{P_0(\vec{\alpha})
H^{\{\mbox{\bf R}\}}_{n_1 \ldots n_N\,n_1 \ldots n_N}\Big(\mbox{\bf y}
(\vec{\alpha}),
\mbox{\bf y}^*(\vec{\alpha})\Big)}{n_1!\cdots n_N!}\geq 0
\end{equation}
for all values of $\mbox{\bf n}$ and $\vec{\alpha}$. In (\ref{eq.s310}) the expressions for
$P_0(\vec{\alpha})$, $\mbox{\bf R}$, and  $\mbox{\bf y}(\vec{\alpha})$
are given by formulas~(\ref{eq.19a})--(\ref{eq.19c}) for photon-number
tomograms.

For Gaussian states, this criterion can be reformulated as
the property of quadrature dispersion matrix $\sigma$.
For example, it has to satisfy the condition 
\begin{equation}\label{eq.s311}
\mbox{det}\pmatrix{\sigma_{p_k p_k}&\sigma_{p_k q_k}\cr
\sigma_{p_k q_k}&\sigma_{q_k q_k}}\geq \frac{1}{4}
\end{equation}
for all $k=1,2,\ldots ,N$. 
The Schr\"odinger--Robertson uncertainty relation
\begin{equation}\label{eq.s312}
\mbox{det}\sigma\geq \left(\frac{1}{4}\right)^N 
\end{equation}
(here $\hbar=1)$  is also necessary condition of positivity of
the Gaussian state. But it is not sufficient condition. 
The necessary and sufficient condition of positivity of the Gaussian state
is the set of inequalities~(\ref{eq.s310}), (\ref{eq.s311}), and
partial Schr\"odinger uncertainty relations for all modes. 
The set of inequalities~(\ref{eq.s311}) for non-Gaussian states is necessary
condition of positivity (nonnegativity) of an Hermitian operator
but they are insufficient for nonnegativity.
For one-mode Gaussian state, the condition of nonnegativity of the density
operator reads 
\begin{equation}\label{eq.s313}
\sigma_{p p}\sigma_{q q} - \sigma_{p q}^2\geq \frac{1}{4},
\end{equation}
which is nothing else as the standard Schr\"odinger--Robertson uncertainty
relation. If this inequality is not fulfilled, one has nonpositive density operators.
But such operator (or such corresponding Wigner function) can describe
the classical state with Gaussian distribution function on the phase space.
Thus negative probabilities related to ``density operators'' are appropriate
to describe the classical states in the standard classical statistical
mechanics. The connection of uncertainty relation for quadratures
(position and momentum) with photon statistics was found in~\cite{23}.

\section{Conclusions }
We have shown that the photon-number tomogram of a generic Gaussian
(one-mode and multimode) state is expressed in terms of multivariable 
Hermite polynomials. This tomogram coincides with the photon-number
distribution function related to the Wigner function with displaced
arguments. The tomogram obtained can be used for measuring quantum states. 
We found the criterion of positivity of the density matrix and consider
this criterion for the photon-number tomograms. 
We discovered necessary and sufficient condition of positivity of the
density matrix in the case of a Gaussian state. We constructed the set of inequalities which is necessary condition
of positivity of the density operator in the case of non-Gaussian states.

\section{Acknowledgments}
The study was partially supported by the Russian Foundation for 
Basic Research under Projects~Nos.~01-02-17745 and 03-02-16408.


\begin{thebibliography}{99}                         

\bibitem{vogel}
S.~Wallentowitz and W.~Vogel, {\sl Phys. Rev.} A, {\bf 53}, 4528 (1996).

\bibitem{wodk}
K.~Banaszek and K.~Wodkiewicz, {\sl Phys. Rev. Lett.}, {\bf 76},
4344 (1996).

\bibitem{euroPLMancini}
S.~Mancini, V. I.~Man'ko, and P.~Tombesi, {\sl Europhys. Lett.}, {\bf 37},
79 (1997).

\bibitem{bertrand}
J.~Bertrand and P.~Bertrand, {\sl Found. Phys.}, {\bf 17}, 397 (1987).

\bibitem{vogelrisken}
K.~Vogel and H.~Risken, {\sl Phys. Rev.} A, {\bf 40}, 2847 (1989).

\bibitem{tombesisemopt95}
S.~Mancini, V.~I.~Man'ko, and P.~Tombesi, {\sl Quantum Semiclass. Opt.},
{\bf 7}, 615 (1995).

\bibitem{jmodopt}
S.~Mancini, V.~I.~Man'ko, and P.~Tombesi, {\sl J. Mod. Opt.}, {\bf 44},
2281 (1997).

\bibitem{dariano}
G.~M.~D'Ariano, S.~Mancini, V.~I.~Man'ko, and P.~Tombesi, {\sl Quantum
Semiclass. Opt.}, {\bf 8}, 1017 (1996).

\bibitem{physlet}
V.~V.~Dodonov and V.~I.~Man'ko, {\sl Phys. Lett.} A, {\bf 239}, 335 (1997)

\bibitem{jetp}
V.~I.~Man'ko and O.~V.~Man'ko, {\sl JETP}, {\bf 85}, 430 (1997).

\bibitem{36}
Olga~ Man'ko and V.~I.~Man'ko, {\sl J. Russ. Laser Res.}, {\bf 18}, 407
(1997).

\bibitem{andreev}
V.~V.~Andreev and V.~I.~Man'ko,  {\sl JETP}, {\bf 87}, 239 (1998). 

\bibitem{safonov}
O.~V.~Man'ko, V.~I.~Man'ko, and S.~S.~Safonov, {\sl Theor. Math. Phys.}, 
{\bf 115}, 185 (1998).

\bibitem{klimov} 
A.~B.~Klimov, O.~V.~Man'ko, V.~I.~Man'ko, Yu.~F.~Smirnov, and V.~N.~Tolstoy, 
{\sl J. Phys. A: Math. Gen.}, {\bf 35}, 6101 (2002). 

\bibitem{castanos}
O.~Castanos, R.~Lopes-Pena, M.~A.~Man'ko, and V.~I.~Man'ko, {\sl J. Phys.
A: Math. Gen.}, {\bf 36}, 4677 (2003); {\sl J. Opt. B: Quantum Semiclass. Opt.},
{\bf 5}, 227 (2003).

\bibitem{cahillglauber}
K.~E.~Cahill and R.~J.~Glauber, {\sl Phys. Rev.}, {\bf 177}, 1882 (1969).

\bibitem{21}
V.~V.~Dodonov, O.~V.~Man'ko, and V.~I.~Man'ko, {\sl Phys. Rev.} A, {\bf 49}, 
2993 (1994). 

\bibitem{vourdas}
A.~Vourdas, {\sl Phys. Rev.} A, {\bf 34}, 3466 (1986).

\bibitem{vourdasweiner}
A.~Vourdas and R.~M.~Weiner, {\sl Phys. Rev.} A, {\bf 36}, 5866 (1987). 

\bibitem{schleih}
W.~Schleich and J.~A.~Wheeler, {\sl J. Opt. Soc. Am.} B, {\bf 4}, 1715 (1987).

\bibitem{marmo} 
O.~V.~Man'ko, V.~I.~Man'ko, and G.~Marmo, {\sl J. Phys. A: Math. Gen.},
{\bf 35}, 699 (2002).

\bibitem{andreevsafonov}
V.~A.~Andreev, O.~V.~Man'ko, V.~I.~Man'ko, and S.~S.~Safonov, 
{\sl J. Russ. Laser Res.}, {\bf 19}, 340 (1998). 

\bibitem{23}
V.~V.~Dodonov, O.~V.~Man'ko, and V.~I.~Man'ko, {\sl Phys. Rev.} A, {\bf 50},
813 (1994).

\bibitem{raymer} D.~T.~Smithey, M.~Beck, M.~G.~Raymer, and A.~Faridani,
{\sl Phys. Rev. Lett.}, {\bf 70}, 1244 (1993).

\end{thebibliography}
\end{document}